\documentstyle[12pt]{article}
\begin{document}
\newcommand{\D}{\mathop{\rm d}}
\title{Error threshold in simple landscapes}
\author{Silvio Franz\\
ICTP, Strada Costiera 10, I-34100 Trieste (Italy)\\
\\
and\\
\\
Luca Peliti\\
Groupe de Physico-Chimie Th\'eorique, URA 1382, ESPCI\\
10, rue Vauquelin, F-75231 Paris Cedex 05 (France)\\
and\\
Dipartimento di Scienze Fisiche and Unit\`a INFM\\
Mostra d'Oltremare, Pad.~19, I-80125 Napoli (Italy)}
\maketitle
\begin{abstract}
We consider the quasispecies description
of a population evolving in
both the ``master sequence'' landscape (where a single sequence
is evolutionarily preferred over all others) and the REM
landscape (where the fitness of different sequences is an
indipendent, identically distributed, random variable).
We show that, in both cases, the error threshold
is analogous to a {\sl first order\/} thermodynamical
transition, where
the overlap between the average genotype
and the optimal one drops discontinuously to zero.
\end{abstract}
An equation describing the behavior of populations of 
self reproducing entities,
subject to natural selection and to mutations, was introduced by Manfred
Eigen in 1971~\cite{Eigen}. The inheritable structure 
(``genotype'') of these entities
is described by a sequence of length $L$ of symbols 
belonging to an alphabet
of $\kappa$ characters ($\kappa=4$ in the case of nucleic acids).
In the simple case in which one such sequence is selectively preferred
with respect to all others, Eigen was able to show that his equation
(the {\sl quasispecies equation\/}) implies
a transition (called the {\sl error threshold\/})
between two different behaviors:
\begin{itemize}
\item At low mutation rate, the population is made up, at equilibrium,
of sequences close to the preferred one ({\sl master sequence\/}):
it forms therefore a {\sl quasispecies\/};
\item At higher mutation rate, the distribution becomes uniform over
sequence space.
\end{itemize}
This behavior is reminiscent of a phase transition in statistical mechanics.
Indeed, I.~Leuth\"ausser~\cite{Leuth} showed that the quasispecies equation 
is equivalent to a statistical mechanical model.
The error threshold corresponds in this language to a thermodynamical
transition of the statistical mechanical system.

Later, Tarazona~\cite{Tara} qualified this correspondence, by pointing out
that the properties which described the behavior of the evolving population
corresponded to {\sl surface\/} observables 
of the statistical mechanical model.
In particular he argued that in the simple situation mentioned above,
with a single master sequence, while the naive application of statistical
mechanics predicted a first-order phase transition, the transition
was continuous for the evolutionary model. The discrepancy between the two
predictions was attributed to a surface phenomenon akin to 
wetting~\cite{Dietrich}, where the disordered state 
is favored on a surface layer
whose thickness diverges as the phase transition is approached.

The statistical mechanics approach to the quasispecies equation was later
used by Franz et al.~\cite{FPS} 
to solve it in a ``rugged fitness landscape''
(in which the selective value of each different sequence is an independent
random variable)
modelled by Derrida's Random Energy Model (REM)~\cite{Derrida}.
In this case, a first-order transition between the quasispecies and the
uniform behavior was found. This result has been challenged by
P. G. Higgs and G. Woodcock~\cite{Higgs96}.

The aim of this letter is to point out that 
the discrepancy between predictions
is due to the fact that one's attention is directed towards different
observables in the different cases: a careful 
consideration of the ``infinite
genome'' ($L\to \infty$) limit, necessary to 
obtain a sharp phase transition, shows
that, in the ``master sequence'' model,  
the error threshold is a first-order
phase transition. This does not rule out
the fact that, in the same limit, the fraction of 
individuals whose genotype
is equal to the master sequence (or, for that matter, 
is at any finite Hamming distance
away from it) goes smoothly to zero at the transition. In particular, the
``wetting phenomenon'' described by Tarazona does not take place, at least
in this case. Similar results hold for the rugged fitness landscape.

Let us consider the $\kappa{=}2$ ``master sequence'' model, 
defined as follows.
The genotype $s$ is described by $L$ units $s_i=\pm 1$, $i=1,\ldots,L$.
The quasispecies equation, which describe the evolution of
the fraction $x_s(t)$ of individuals having 
the genotype $s$ at generation $t$,
takes the form
\begin{equation}
x_s(t+1)=\frac{1}{Z(t)}\sum_{s'}Q_{ss'}w_{s'}x_s(t),\label{QS}
\end{equation}
where $w_s$ is the fitness of sequence $s$ and
$\|Q_{ss'}\|$ is the mutation matrix.
The normalization factor $Z(t)$ is given by
\[Z(t)=\sum_sw_s x_s(t).\]
The matrix element $Q_{ss'}$ is the conditional probability
that a reproduction event of an individual with genotype $s'$
produces one with genotype $s$. If one assumes pointwise mutations
with uniform probability one has
\begin{equation}
Q_{ss'}=\mu^{d_{\rm H}(s,s')}(1-\mu)^{L-d_{\rm H}(s,s')},\label{mutation}
\end{equation}
where 
\begin{equation}d_{\rm H}(s,s')=\frac{1}{2}\sum_{i=1}^L\left(1-s_is'_i\right)
\end{equation}
is the Hamming distance between the sequences $s$ and $s'$, and $\mu$
is the mutation rate.
The ``master sequence'' is denoted by $s^0=(s^0_i)$.
The fitness $w_s$ is then given by
\begin{equation}
w_s=\cases{\exp(kL),&if $s=s^0$;\cr
1,&otherwise.}\label{fitness}
\end{equation}
In this expression, $k>0$ is a ``selective'' inverse temperature.
We have chosen to take $\ln\, w_{s^0}\propto L$ in order to obtain
the infinite genome limit in close analogy
with the thermodynamical limit. We shall discuss later the 
scaling considered by Eigen~\cite{Eigen}
and followers, in which $w_{S^0}\to {\rm const.}$

As pointed out by Leuth\"ausser and Tarazona~\cite{Leuth,Tara},
the solution of eq.~(\ref{QS}) can be expressed in terms of a statistical
mechanical model. Let us consider a population evolving for $T$
generations from an initial condition in which $x_s=\delta_{ss^0}$.
One has
\begin{eqnarray}
x_s(T)&=&\frac{1}{{\cal Z}}\sum_{s(1),s(2),\ldots,s(T-1)}
Q_{s(T)s(T-1)}w_{s(T-1)}\cdots Q_{s(1)s^0}{w_{s^0}}\nonumber\\
&=&\frac{1}{{\cal Z}}\sum_{s(1),s(2),\ldots,s(T-1)}
\exp\left[\sum_{t=1}^T\left(\ln Q_{s(t)s(t-1)}+\ln w_{s(t-1)}\right)\right].
\end{eqnarray}
We have set $s(T)=s$, $s(0)=s^0$, and we have defined the normalization constant
${\cal Z}$ by
\begin{eqnarray*}
{\cal Z}
&=&\sum_{s(1),s(2),\ldots,s(T-1),s(T)}
\exp\left[\sum_{t=1}^T\left(\ln Q_{s(t)s(t-1)}+\ln w_{s(t-1)}\right)\right]\\
&\equiv&\sum_{s(1),s(2),\ldots,s(T-1),s(T)}\exp(-H\{s(t)\}).
\end{eqnarray*}
The last line defines the symbol $H$.
It now turns out that, for the ``master sequence'' model,
\begin{eqnarray}
-H\{s(t)\}&\equiv&\sum_{t=1}^T\left(\ln Q_{s(t)s(t-1)}+\ln\, w_{s(t-1)}\right)\nonumber\\
&=&TL\ln\,(1-\mu)\nonumber\\
&&+\sum_{t=1}^T\left(\beta\sum_{i=1}^L s_i(t)s_i(t-1)+k\delta_{s(t-1)s^0}\right),\label{Hamilton}
\end{eqnarray}
where the ``mutation'' inverse temperature $\beta$ is defined by
\begin{equation}
\beta=\frac{1}{2}\ln\frac{1-\mu}{\mu}.
\end{equation}
The expression (\ref{Hamilton}) looks like the Hamiltonian (times the temperature)
of an Ising system of $TL$ spins, arranged in $T$ layers of $L$ spins each.
The interlayer interactions, representing the correlation effects due to the
heredity, are proportional to $\beta$, while the intralayer interactions,
representing the selection, are proportional to $k$. Tarazona~\cite{Tara}
pointed out that the intralayer interaction term corresponding
to layer $T$ is lacking in this expression: the system corresponds
therefore to a statistical mechanical model with a free surface.

It is now easy to see that, in the limit $L\to\infty$ followed by $T\to\infty$,
a phase transition separates an ordered (``frozen'') regime in which 
one has $s(t)=s^0$ for all layers $t$ except
the last one, from a disordered (``free'') one, in which all sequences $s$ have the same
probability, and the system behaves like a collection of $L$ independent
one-dimensional Ising models at temperature $\beta^{-1}$.
The transition line can be obtained by comparing the free energies $F$
defined by $F=-\ln\,{\cal Z}$:
\begin{enumerate}
\item For the ordered regime one has
\begin{equation}
F_1=-TL\,(k+\beta)+{\rm \ boundary\ terms};
\end{equation}
\item For the disordered regime one has
\begin{equation}
F_2=-TL\,\ln( 2\cosh\beta)+{\rm\ boundary\ terms},
\end{equation}
corresponding to the free energy per spin of a one-dimensional Ising model.
\end{enumerate}
The transition line is given by the condition $F_1=F_2$ (where the boundary terms
are neglected) and reads
\begin{equation}
k_{\rm t}(\beta)=\ln (2\cosh\beta)-\beta.
\end{equation}
In terms of the mutation rate $\mu$, this corresponds to
$k_{\rm t}=|\ln(1-\mu)|$, as originally obtained by Eigen~\cite{Eigen}.

We now show in more detail that all layers but the last one
(corresponding to $t=T$) are ``frozen"
for $k>k_{\rm t}(\beta)$,
in the sense that the only configurations which contribute in the
infinite genome limit are those for which $s(t)=s^0$ for $t<T$.
Let us consider the last-but-one layer ($t=T-1$), and let us momentarily
assume that the preceding layer is frozen. The last layer is
free, because there are no contributions from the
intralayer interactions at $t=T$~\cite{Tara}.
There are two possibilities for $s(T-1)$:
\begin{enumerate}
\item ``Frozen": $s(T-1)=s^0$: this yields a contribution
$\exp\left[L(k+\beta)\right]\times(2\cosh\beta)^L$ to the partition sum;
the second factor comes from the sum over the configurations of the
last layer;
\item ``Free'': summing also over the configurations of the last-but-one layer,
one obtains the contribution $(2\cosh\beta)^{2L}$.
\end{enumerate}
Because one has, by hypothesis, $k>k_{\rm t}(\beta)=\ln (2\cosh\beta)-\beta$,
the first contribution dominates for $L\to\infty$.
By induction, one can show in the same way that it is not possible
that there is a label $t_0<T$ separating ``frozen'' layers (for $t<t_0$)
from ``free'' ones (for $t\ge t_0$).

Let us now define, following Tarazona~\cite{Tara}, the order parameter
$m$ as the overlap of the average sequence $\left(\left<s_i\right>\right)$
with the master sequence $s^0$:
\begin{equation}
m=\frac{1}{L}\sum_{i=1}^L\left<s_i\right>s^0_i=1-2\left<d_{\rm H}(s,s^0)\right>/L.
\end{equation}
The angular brackets denote the population average:
\begin{equation}
\left<A(s)\right>=\sum_sx_s A(s),
\end{equation}
where we have taken into account that $\sum_s x_x=1$.

In the ordered phase, all layers but the last one are frozen to the master sequence.
It is then a simple matter to show that
\begin{equation}
m=\tanh \beta=1-2\mu.\label{ordpar}
\end{equation}
On the other hand, $m=0$, obviously, in the disordered phase.
We have thus obtained the result that the phase transition is of {\sl first order},
and that $m$ drops discontinuously from $1-2\mu$ to 0 as $k$ falls
below the transition value $k_{\rm t}$.
Let us also remark that eq.~(\ref{ordpar}) predicts that $m=0$ for $\beta=0$,
even for $k>k_{\rm t}(0)=\ln 2$, as it is reasonable to expect on intuitive grounds.

This analysis is supported by the numerical solution of the quasispecies
equation for finite $L$. We show in fig.~1 the order parameter as a function of
$\mu$ for different values of $L$. The value of $k$ is such that the error threshold
takes place for $\mu=\mu_{\rm t}=0.25$.
One  clearly sees that the curve approaches a discontinuous behavior
as $L$ increases, contrary to the statements contained in ref.~\cite{Tara}.
Let us remark that, properly speaking, the weight $x_{s^0}(T)$
in the population approaches 0 in the thermodynamical limit ($L\to\infty$,
$k,\beta={\rm const.}$). Nevertheless the population forms
a {\sl bona fide\/} quasispecies, in the sense of ref.~\cite{Eigen}.

Eigen~\cite{Eigen}, Leuth\"ausser~\cite{Leuth}, and Tarazona~\cite{Tara}
have considered a situation
in which the fitness ratio $w_{s^0}/w_s$ is kept constant as $L$ increases.
In this case, if the mutation rate $\mu$ (and hence $\beta$) is kept constant,
one eventually crosses over smoothly to a ``disordered'' regime, independently
of the value of this ratio. This is the point that Eigen wanted to make
when he introduced the quasispecies equation, back in 1971: that the error threshold
prevented biological information to be maintained, if genome length
exceeded a certain value.

In this situation, there is no sharp phase transition, and the question whether
it be of first or second order is pointless. However, even in this case, one can obtain
a phase transition in the limit $L\to\infty$, if one mantains constant
the average number $\mu L$ of mutations. This corresponds to take
$\ln\,\beta\propto L$. It is possible to solve the problem in this limit,
and the results concide with what one obtains by taking the same limit
in the equations we haver written above. In particular, the transition
is still of first order, but now the order parameter $m$ jumps from 1 to 0:
just above the transition, the {\sl whole\/} population lies a finite Hamming
distance away from the master sequence (even though the weight of the
master sequence goes smoothly to zero).

On the other hand, this behavior does contradict the fact that
the weight $x_{s^0}(T)$ of the master sequence (which does not vanish
if $\mu$ is small enough) approaches 0 {\sl continuously\/}
at the error threshold, as exhibited by fig.~2. 
The limit behavior is indeed given
by~\cite{Higgs93}
\begin{equation}
x_{s^0}=\cases{1-u/u_{\rm t},&for $u<u_{\rm t}$;\cr
0,&otherwise;}
\end{equation}
where $u=1-\exp(-\mu L)$ is the total mutation probability,
and $u_{\rm t}$ is the corresponding transition value.
However, even in the infinite genome limit, the whole population is the offspring
of master sequence individuals at each generation, and has therefore
a finite overlap with the master sequence, as soon as one is above the transition.

Let us now consider the REM fitness landscape. In this case the
fitness $w_s$ is given by
\begin{equation}
w_s=\exp\left(-kE(s)\right),
\end{equation}
where the ``energies'' $E(s)$ of different sequences are independent
normally distributed random variables, with zero average
and variance equal to $L/2$. We have correspondingly
\begin{equation}
{\cal Z}=\sum_{s(1),\ldots,s(T)}\exp\left[\sum_{t=1}^{T-1}
\left(\beta\sum_{i=1}^Ls_i(t)s_i(t+1)-kE(s(t))\right)\right],
\end{equation}
where, as before, we assume that $s(0)$ corresponds to an energy minimum.
The bulk properties of this model have been studied in~\cite{FPS} with
the replica method. We briefly illustrate here the results, using
the argument originally developed by Derrida to solve the REM~\cite{Derrida}.
Suppose to consider two neighboring layers, $t$ and $t+1$,
whose overlap $q$, defined by
\[ q=\frac{1}{L}\sum_{i=1}^Ls_i(t)s_i(t+1),\]
has a fixed value. The average number of these configurations
with energy equal to $E$ is given by
\begin{equation}
{\cal N}(E,q)\sim \exp\left(LS(q)-E^2/L\right),\label{nu}
\end{equation}
where
$S(q)=\ln\,2-\frac{1}{2}\left[(1+q)\ln(1+q)+(1-q)\ln(1-q)\right].$
The typical value is equal to the average value if
the latter is exponentially large, and vanishes otherwise.
 We can thus write for a typical sample:
\begin{equation}
{\cal Z}=\int_{(LS(q)-E^2/L)>0}
\D E\D q\,{\cal N}(E,q)\,\exp\left[T(-kE+L\beta q)\right].
\end{equation}
This expression is dominated by the saddle point in the free phase,
and by the smallest value of the energy (and $q=1$) in the frozen phase.
The typical value can be obtained by eq.~(\ref{nu}), by setting
${\cal N}(E,0)\sim O(1)$, and is equal to $-L\sqrt{\ln\,2}$.
The free energy is thus given 
\begin{equation}
F=\cases{-TL\left[\ln(2\cosh\beta)+k^2/4\right],&in the free phase;\cr
-TL\,(k\sqrt{\ln\,2}+\beta),&in the frozen phase.}\label{fre}
\end{equation}
By comparison of the free energies, the transition line is
located at~\cite{FPS}
\begin{equation}
k_{\rm t}(\beta)=2\left(\sqrt{\ln\,2}-\sqrt{\beta-\ln(\cosh \beta)}\right).
\end{equation}
We remark {\sl en passant\/} that, contrary to the REM and other systems
with discontinuous glass transitions~\cite{KT}, here the
transition is thermodynamically of first order, with a latent
heat that can be easily computed from eq.~(\ref{fre}).

The surface (evolutionary) properties can be worked out as in the ``master
sequence'' case. Let us consider the frozen phase,
and let us assume that layer $T-2$ is frozen into one of the REM ground states.
We focus on layer $T-1$ and assume that layer $T$ is free.
Layer $T$ has then no influence on layer $T-1$, whose contribution
to the free energy is given by
\begin{equation}
{\cal Z}_{T-1}=\sum_{s(T-1)}\exp\left(-kE(s(T-1))+\beta\sum_is_i(T-2)
s_i(T-1)\right).\label{lyer}
\end{equation}
Analyzing eq.~(\ref{lyer}) as above, we find a freezing transition
into the ground state, with the same conditions as for the bulk.
Therefore, as long as $k>k_{\rm t}$, all layers but the last one
are frozen in the energy ground state. It is clear at this point
that all along the frozen phase $m=(1/L)\sum_i\left<s_i(0)s_i(t)\right>$
is independent of $t$ for $t>0$, and is given by $m=\tanh\beta$.
In the same way the weight of the optimal sequence behaves as
in the ``master sequence'' model.

Summarizing, we have shown that in the ``master sequence'' and in the
REM landscapes for the quasispecies equation, in the limit in which one
can speak of sharp phase transitions, surface phenomena do not appear.
Indeed the surface passively follows the behavior of the bulk. This is
due to the pathology of the model, that is one-dimensional in the time
direction, but mean-field like in the sequence direction.
The analysis also shows that ``master sequence'' and REM landscapes, apart
from details, have very similar evolutive properties.

\section*{Acknowledgments}
SF thanks the Laboratoire de Physique Th\'eorique
of the Ecole Normale Sup\'erieure (Paris) for hospitality and
support during the elaboration of this work.
LP is Associato to the Istituto Nazionale di Fisica Nucleare, Sezione di Napoli,
and acknowledges the support of a Chaire Joliot of the ESPCI.

\section*{Figure captions}
\begin{enumerate}
\item Order parameter $m$ in the ``master sequence'' model as a function
of the mutation rate $\mu$ for $L=10,20,40,80$. The selective
temperature $k$ equals $\ln\,(4/3)$.
Also plotted the prediction of eq.~(\ref{ordpar}).
\item Weight $x_{s^0}=x_0$ of the master sequence as a function of the total
mutation rate $u=1-\exp(-\mu L)$ for $L=10,20,40,80$.
We have chosen $kL=1/4$ so that $\mu_{\rm t}L=1/4$.
Also plotted the prediction~\cite{Higgs93}
$x_{s^0}=1-u/u_{\rm t}$, where $u_{\rm t}$ is the value 
of $u$ at the transition.
\end{enumerate}

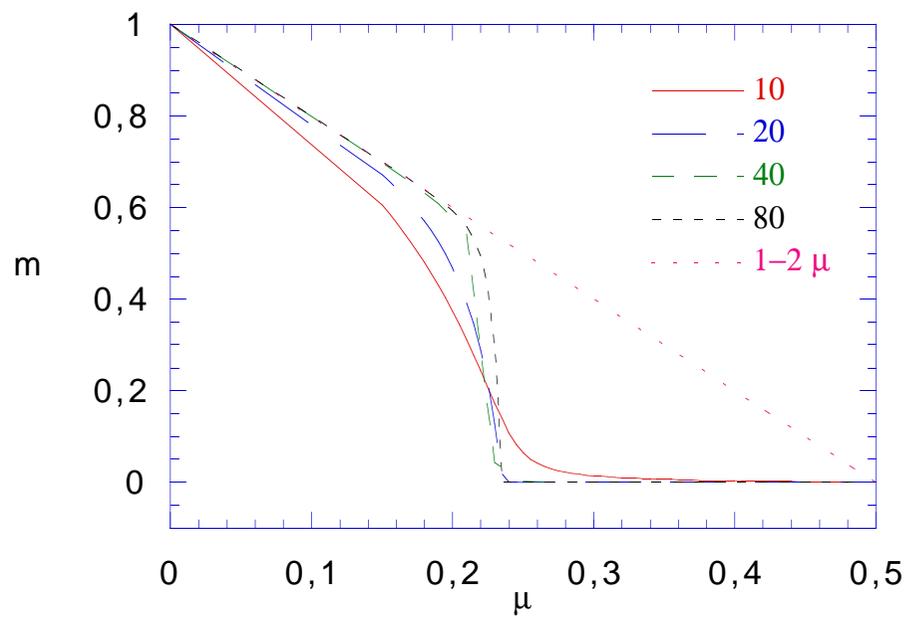
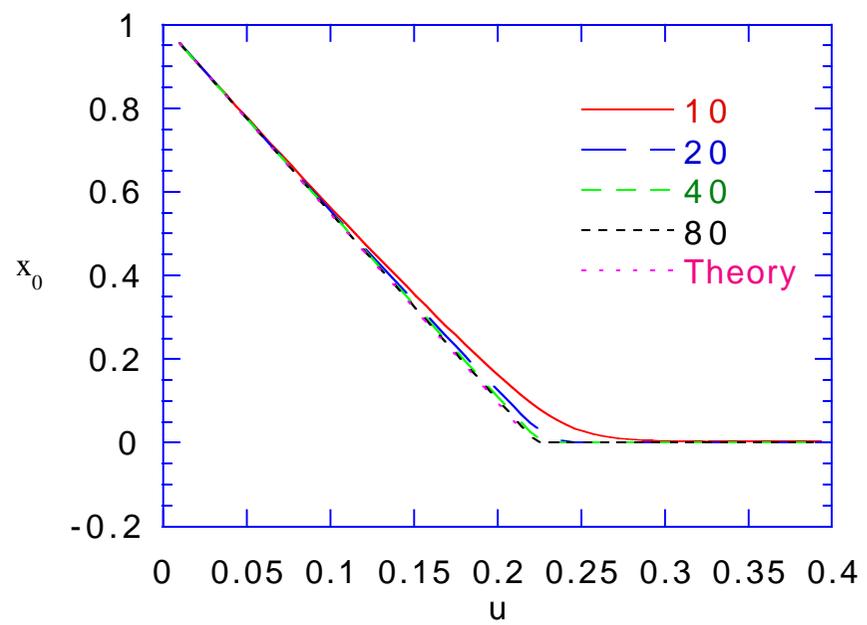
\end{document}